\documentclass[cameraready]{Interspeech}
\usepackage{amsmath,graphicx,hyperref}
\usepackage{arydshln}
\usepackage{multirow}

\usepackage{booktabs}
\usepackage{tabularx}
\title{WildElder: A Chinese Elderly Speech Dataset from the Wild with Fine-Grained Manual Annotations}

\author[affiliation={1}, orcid=0009-0003-8057-4644, equalcontribution]{Hui}{Wang}
\author[affiliation={1}, orcid=0009-0002-4819-4572, equalcontribution]{Jiaming}{Zhou}
\author[affiliation={1}, orcid=0000-0001-6448-7784]{Jiabei}{He}
\author[affiliation={1}, orcid=0000-0002-8554-8969]{Haoqin}{Sun}
\author[affiliation={1,2}, orcid=0009-0000-2748-3020, correspondingauthor]{Yong}{Qin}

\address{
    $^1$ College of Computer Science, Nankai University, Tianjin, China \\ 
    $^2$ Academy for Advanced Interdisciplinary Studies, Nankai University, Tianjin, China
}

\email{wanghui\_hlt@mail.nankai.edu.cn, qinyong@nankai.edu.cn}

\keywords{speech recognition, elderly speech, in-the-wild, mandarin dataset}

\usepackage{comment}

\begin{document}

\maketitle

\begin{abstract}
    Elderly speech poses unique challenges for automatic processing due to age-related changes such as slower articulation and vocal tremors. Existing Chinese datasets are mostly recorded in controlled environments, limiting their diversity and real-world applicability. To address this gap, we present WildElder, a Mandarin elderly speech corpus collected from online videos and enriched with fine-grained manual annotations, including transcription, speaker age, gender, and accent strength. Combining the realism of in-the-wild data with expert curation, WildElder enables robust research on automatic speech recognition and speaker profiling. Experimental results reveal both the difficulties of elderly speech recognition and the potential of WildElder as a challenging new benchmark. The dataset and code have been available at \url{https://github.com/NKU-HLT/WildElder}.
\end{abstract}

\section{Introduction}

Elderly speech has drawn increasing attention in recent years as aging populations worldwide create growing demands for inclusive speech technologies \cite{scuteri2024aging}. Compared with younger speakers, elderly speech often exhibits reduced volume, slower articulation, tremors, and other age-related changes, which present unique challenges for automatic processing \cite{bona2014temporal,ivanova2024speech}. Robust speech technologies for seniors are of great importance in applications such as voice assistants, healthcare monitoring, and human–computer interaction in daily life. However, research on elderly speech remains hindered by the scarcity of representative and diverse datasets~\cite{10097275,10286152,hu2024self}.

In the context of Chinese, several corpora have been developed to capture the speech characteristics of elderly speakers, yet notable limitations remain. AISHELL-ASR0060 provides read speech from speakers aged 55 and above, yet only includes scripted utterances. The Mandarin Elderly Cognitive Speech Database (MECSD)~\cite{mecsd} is designed for cognitive health research, but its controlled tasks and recording environment limit naturalness. SeniorTalk~\cite{seniortalk} focuses on conversational speech, but it is also collected in structured settings with recruited participants. As a result, most existing datasets are gathered under controlled or laboratory conditions. This limitation leads to insufficient topical diversity, reduced spontaneity in speaking styles, and a lack of variability in recording conditions, which in turn diminishes their utility for developing speech technologies that must function reliably in real-world scenarios.


In parallel, a series of works such as WenetSpeech~\cite{wenetspeech}, Emilia~\cite{emilia}, and GigaSpeech~\cite{gigaspeech,gigaspeech2} demonstrate the feasibility of collecting speech from the wild. These datasets show that harvesting data from online platforms and applying automated pipelines for segmentation, transcription, and filtering can yield corpora with rich topics and acoustic diversity~\cite{wenetspeechyue}. However, their reliance on automated tools inevitably introduces errors, and fine-grained speaker attributes are rarely available~\cite{kuhn2024measuring}. More importantly, the performance of current automatic processing tools is particularly weak on elderly voices, making it even harder to ensure annotation quality and reliability.

\begin{figure}[t]
  \centering
  \includegraphics[width=\linewidth]{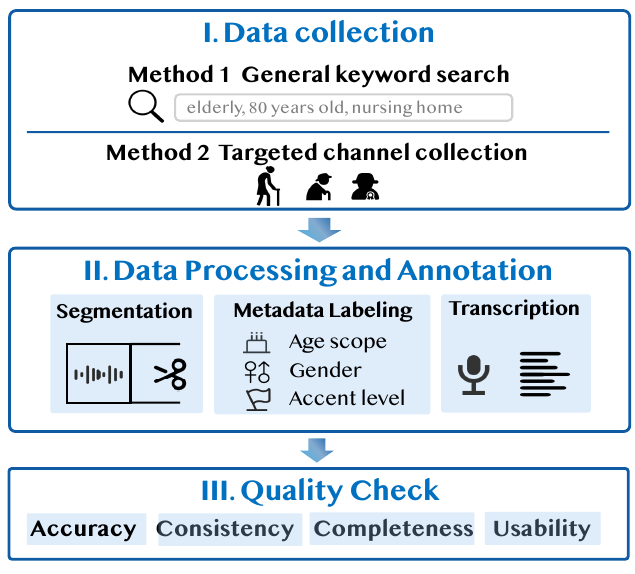}
  \caption{Workflow of the WildElder dataset construction, including data collection, annotation, and quality check.}
  \label{fig:workflow}
\end{figure}


Motivated by these gaps, we introduce WildElder, a Chinese elderly speech dataset collected from the wild and enriched with fine-grained manual annotations. Through the data collection and annotation process shown in Figure~\ref{fig:workflow}, WildElder combines the diversity and realism of in-the-wild data sources with the reliability of expert manual labeling and quality checks. Each utterance is annotated with orthographic transcripts, demographic attributes such as age group and gender, and degree of accent strength. Beyond data construction, we further conduct extensive experiments and analyses on WildElder, evaluating different models and demographic conditions. The results highlight both the difficulty of elderly speech processing and the potential research directions enabled by this dataset, establishing WildElder as a reliable and challenging benchmark for future work. Our contributions are summarized as follows:

\begin{itemize}
  \item We construct \textbf{WildElder}, a real-world Mandarin speech dataset of elderly speakers collected from online videos, providing more diverse and challenging audio compared with traditional laboratory-recorded datasets.
  \item We manually segment and annotate the data with multiple labels, including transcription, speaker age, gender, and accent intensity, enabling research on both ASR and auxiliary tasks such as speaker profiling.
  \item We perform extensive experiments on WildElder and analyze performance across different demographic conditions, demonstrating the challenges of elderly speech in the wild.
\end{itemize}

\begin{table}[t]
\centering
\caption{Accent-strength annotation rubric used in WildElder.}
\label{tab:accent_strength}
\small
\setlength{\tabcolsep}{6pt}
\renewcommand{\arraystretch}{1.2}

\begin{tabularx}{\linewidth}{@{}lX@{}}
\toprule
\textbf{Level} & \textbf{Guideline} \\
\midrule
\textbf{Light} &
Near-standard Mandarin with occasional regional traces; intelligibility is typically unaffected. \\
\addlinespace[2pt]
\textbf{Moderate} &
Noticeable regional pronunciation/prosodic patterns; generally understandable but may require careful listening or an occasional replay. \\
\addlinespace[2pt]
\textbf{Heavy} &
Strong, pervasive regional influence with frequent departures from standard Mandarin; comprehension often benefits from contextual information. \\
\bottomrule
\end{tabularx}

\end{table}

\section{Dataset Description}

\subsection{Construction Process}
As illustrated in Figure 1, the construction of WildElder follows a three-step pipeline covering data collection, processing with manual annotation, and rigorous quality checks. This design allows us to balance the natural diversity of in-the-wild speech with the reliability of curated datasets.

\subsubsection{Data Collection} 
Collecting speech data from elderly speakers is particularly challenging, as such resources are scarce and not systematically available online. To address this, we employ two complementary strategies. The first involves constructing our own keyword list (e.g., “80 years old,” “nursing home”) to perform broad searches across online platforms, allowing us to capture a wide range of speech samples. The second focuses on identifying channels featuring elderly content creators, which provides a more efficient way to collect data with relatively higher quality and reliability. By combining these strategies, we are able to significantly expand and diversify the dataset, ultimately obtaining 619 videos with a total duration of over 71 hours, covering varied scenarios and speaking styles.

\subsubsection{Data Processing and Annotation} 
The raw audio--video materials are manually segmented into utterance-level clips based on sentence boundaries and speaker turns. We remove segments containing non-elderly speakers, excessive background music, overlapping speech, or severe acoustic distortions to retain intelligible elderly speech.

Each retained utterance is transcribed into orthographic Chinese characters under a unified protocol. We preserve natural disfluencies such as repetitions/stuttering, avoid pronunciation normalization for accented speech, standardize numbers (including decimals and digit strings) into Chinese uppercase numerals, and write alphabetic characters in uppercase ASCII. Clips shorter than three Chinese characters are excluded.

Annotators additionally provide metadata including speaker age group, gender, and accent strength. Age group is estimated by triangulating external contextual cues (e.g., title/description, on-screen text, self-introductions, and uploader/channel profile) with visual and auditory evidence. Ambiguous cases are flagged for secondary review. Accent strength is annotated following the rubric in Table~\ref{tab:accent_strength}.

\begin{figure}[t]
  \centering
  \includegraphics[width=\linewidth]{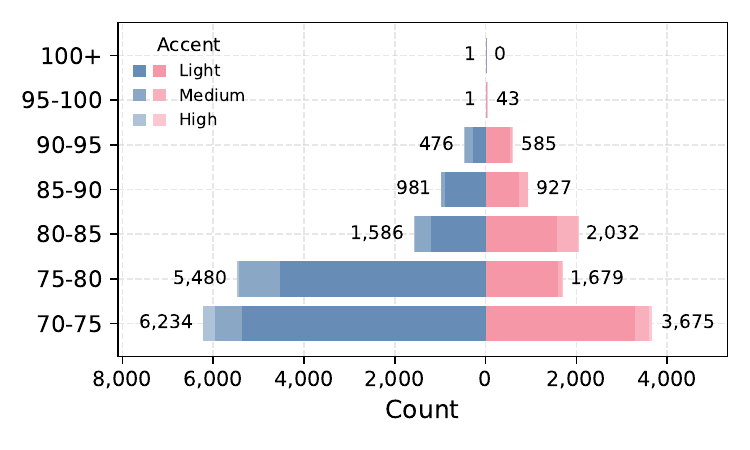}
  \caption{Utterance-level distribution across age groups, separated by gender (male on the left, female on the right) and accent strength (light, medium, heavy).}
  \label{fig:pyramid}
\end{figure}

\subsubsection{Quality Check}
We conduct quality checks along four dimensions: accuracy, completeness, consistency, and usability. 
For \textit{accuracy}, we randomly sample utterances and manually compare the transcripts and metadata labels with the corresponding audio. A sample is counted as correct only if the transcript matches the spoken content and the associated labels (e.g., age group, gender, and accent strength) are judged reasonable given the available evidence. We require an accuracy rate of at least 95\% on the audited samples; batches that fail to meet this threshold are sent back for correction and re-checking.

For \textit{completeness}, we verify that every utterance is paired with valid audio and non-empty annotations, and that all required fields are present without corruption. 
For \textit{consistency}, we inspect samples for adherence to the transcription and labeling guidelines (e.g., number normalization, punctuation conventions, and accent-strength rubric), and resolve ambiguous cases through secondary review. 
For \textit{usability}, we validate the dataset format and specifications, including file integrity, sampling rate/encoding compliance, and segment boundaries. Only utterances that satisfy all criteria are retained in the final release.
\begin{figure}[t]
  \centering
  \includegraphics[width=\linewidth]{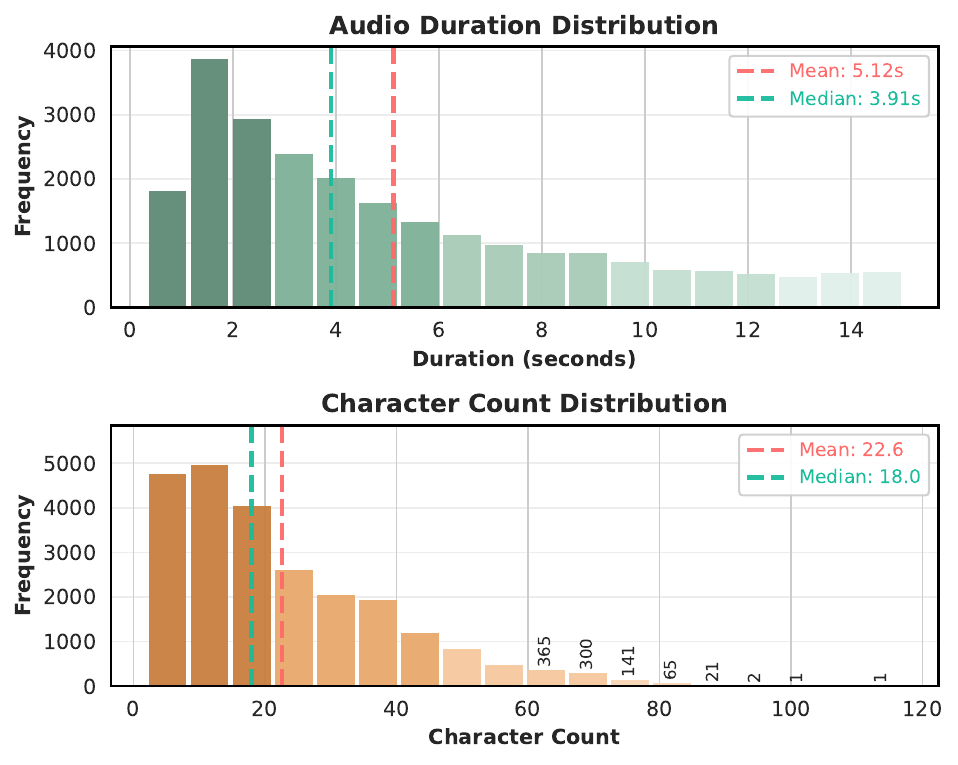}
  \caption{Histograms of utterance duration (top, seconds) and transcript character count (bottom). Dashed lines mark the mean (red) and median (cyan)}
  \label{fig:duration}
\end{figure}

\begin{figure}[t]
  \centering
  \includegraphics[width=\linewidth]{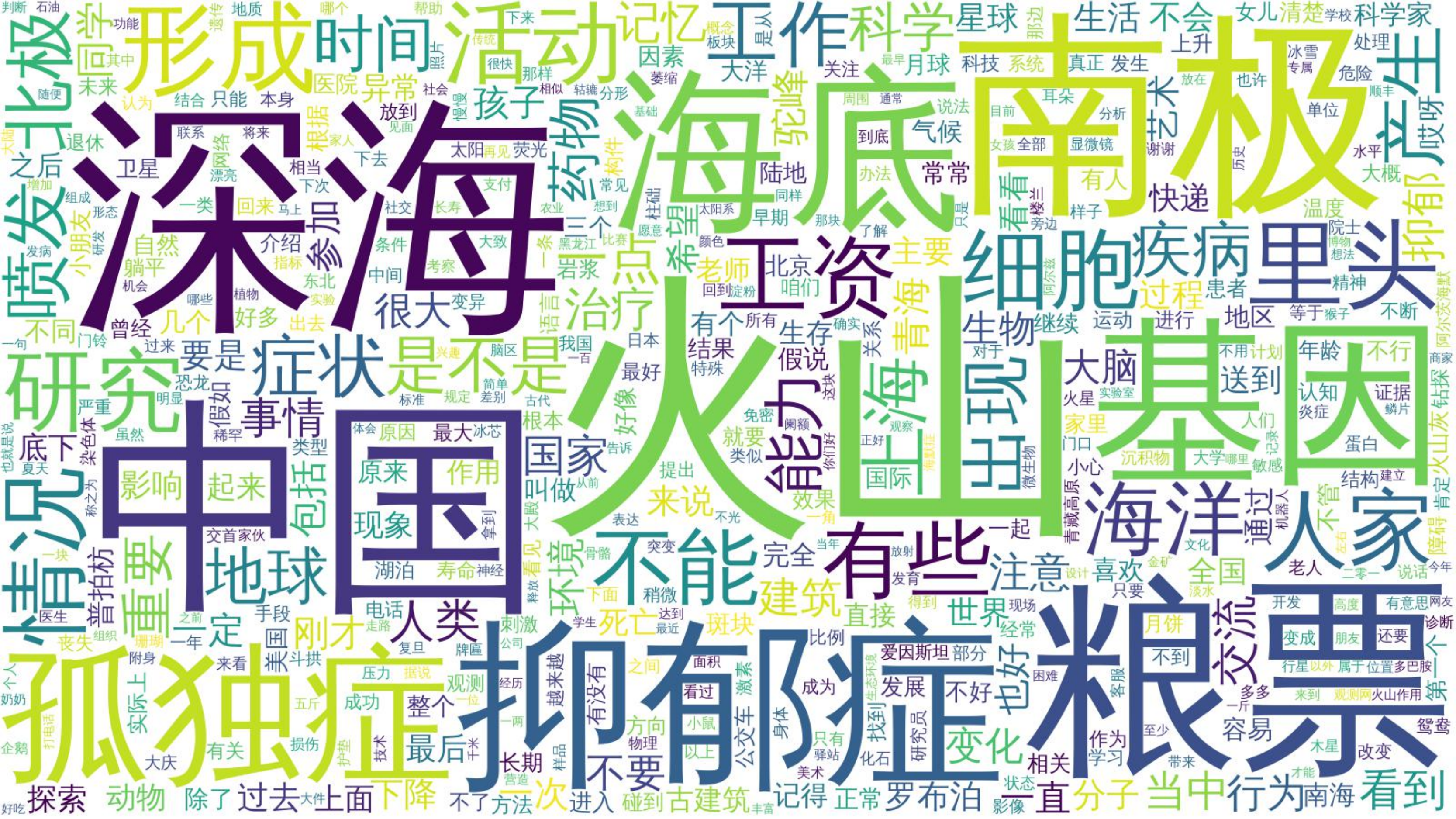}
  \caption{Word cloud of the dataset transcripts, showing frequently used characters and topics.}
  \label{fig:wordcloud}
\end{figure}

\subsection{Dataset Statistics}

The final release of WildElder consists of 23,701 utterances with a total duration of 33.7 hours, collected from 619 videos. The speech is in Mandarin Chinese, with some utterances carrying regional accents. Each utterance is annotated with transcription, speaker age group, gender, and accent strength. The dataset covers a wide range of general topics, including family and daily life, nature and environment, science and education, history and culture, as well as health and disease.

\subsubsection{Demographic Diversity}

Figure~\ref{fig:pyramid} presents the utterance-level distribution across age, gender, and accent strength. The majority of utterances come from speakers between 70 and 85 years old, while speech from individuals above 90 is also included, albeit in smaller amounts. Both male and female speakers contribute substantially, ensuring that the dataset captures gender variation. In terms of accent, light and medium levels dominate, with fewer heavily accented utterances. Overall, this distribution reflects the natural availability of elderly speech in online sources while still covering multiple demographic dimensions, making the dataset suitable for analyzing how age, gender, and accent jointly influence speech characteristics.

\subsubsection{Acoustic and Linguistic Richness}
Figure~\ref{fig:duration} summarizes the acoustic and linguistic characteristics of the dataset. Most utterances are short, with a median duration of 3.91 seconds. In terms of textual length, the character count shows a similar pattern, with a median of 18 characters. The long-tail distributions include longer utterances and longer texts, providing additional variability for modeling extended speech segments. 

\begin{table}[t] 
    \centering 
    \caption{Training configurations for different models.} 
    \vspace{2pt} 
    \label{tab:training-config} 
    \begin{tabular}{l c c c} 
        \hline 
        \textbf{Model} & \textbf{Batch Size} & \textbf{Learning Rate} & \textbf{Epochs} \\ 
        \hline 
        Transformer & 32 & $1\times10^{-3}$ & 100 \\
        Conformer & 32 & $1\times10^{-3}$ & 100 \\ Branchformer  & 16 & $1\times10^{-3}$ & 100 \\ 
        Paraformer  & 32 & $1\times10^{-3}$ & 100 \\ 
        CW   & 16 & $4\times10^{-5}$ & 20 \\ 
        Whisper   & 16 & $1\times10^{-5}$ & 20 \\ 
        \hline 
    \end{tabular} 
\end{table}

The word cloud in Figure~\ref{fig:wordcloud} further indicates broad topical coverage (e.g., daily life, health, science, and social issues), underscoring the dataset’s linguistic diversity. In addition to everyday conversations, the transcripts also touch on technology, culture, and mathematics, reflecting a wider range of semantic content compared with past datasets \cite{seniortalk}.

\begin{table*}[t]
\centering
\caption{Character Error Rate (CER, \%) of models trained from scratch under different decoding strategies.}
\vspace{2pt}
\label{tab:asr_results}
\begin{tabular}{lcccccc}
\toprule
\textbf{Model} & \textbf{Loss} & \textbf{Parameter} & \textbf{CTC Greedy} & \textbf{CTC Beam} & \textbf{Attention} & \textbf{Attention Rescoring} \\
\midrule
Transformer  & CTC+ATT & 29.80M & 37.44 & 37.28 & 47.54 & 36.30 \\
Conformer  & CTC+ATT & 31.94M & 32.51 & 32.48 & 39.58 & 31.74 \\
Branchformer  & CTC+ATT & 29.01M & 35.67 & 35.61 & 45.60 & 34.80 \\
Paraformer  & CTC+Paraformer & 31.04M & 42.71 & 38.39 & -     & -     \\
\bottomrule
\end{tabular}

\end{table*}

\begin{table}[htbp]
\centering
\caption{Zero-shot and fine-tune CER (\%) performance.}
\vspace{2pt}
\label{tab:whisper_results}
\begin{tabular}{lcc}
\toprule 
\textbf{Model} & \textbf{Zero-shot} & \textbf{Fine-tune} \\
\midrule 
Conformer-WenetSpeech             & 16.43 & 13.54 \\
Whisper-Tiny    & 53.50 & 32.20 \\
Whisper-Base    & 41.67 & 26.61 \\
Whisper-Small   & 29.00 & 20.25 \\
Whisper-Medium  & 23.41 & 16.14 \\
\bottomrule
\end{tabular}

\end{table}

\section{Experiment Setup}

\subsection{Dataset Splits}
To ensure reproducibility, WildElder is divided into training, development, and test sets at the speaker level. The training set contains 18,835 utterances (26.7 h), the development set 2,465 utterances (3.5 h), and the test set 2,400 utterances (3.5 h). The average utterance duration is around 5.1 seconds across all subsets, showing a balanced partition suitable for training and evaluation.

\subsection{Baseline Models}
We establish two groups of baselines to assess the difficulty of WildElder, Table~\ref{tab:training-config} summarizes the training configurations.

\subsubsection{Models trained from scratch} 
The scratch-trained set includes Transformer, Conformer~\cite{conformer}, Branchformer~\cite{Branchformer}, and Paraformer~\cite{paraformer} architectures. Transformer provides a pure self-attention sequence-to-sequence framework, while Conformer augments it with convolution to capture local acoustic cues. Branchformer employs a branched attention–feedforward design for richer representations, and Paraformer is a non-autoregressive model combining CTC and parallel decoding for faster inference.

\subsubsection{Pre-trained models.} 

The pre-trained group covers Conformer-WenetSpeech (CW) and the Whisper family (Tiny, Base, Small, Medium)~\cite{whisper}. Conformer-WenetSpeech benefits from large-scale WenetSpeech training, and Whisper models, trained on massive multilingual audio–text pairs, support strong zero-shot recognition and can be fine-tuned on WildElder.

\subsection{Evaluation Metrics}
We evaluate recognition performance using the \textit{character error rate} (CER), which is widely adopted for Mandarin ASR. CER is computed as
\begin{equation}
\mathrm{CER} = \frac{S + D + I}{N} \times 100\%,
\end{equation}
where $S$, $D$, and $I$ denote the numbers of substitutions, deletions, and insertions, and $N$ is the number of characters in the reference transcript. 

\section{Results and Analysis}

\subsection{Models trained from scratch}
Table~\ref{tab:asr_results} shows that all systems trained from scratch still yield relatively high CERs on the WildElder test set, highlighting the difficulty of recognizing in-the-wild elderly speech. Although the models have comparable parameter sizes, decoding strategy has a clear and consistent impact on performance. Pure attention decoding produces the highest error rates across architectures, whereas CTC-based decoding (greedy and beam search) is notably more robust.

Moreover, attention rescoring achieves the best CER for Transformer, Conformer, and Branchformer, outperforming their corresponding CTC-only decoding results. This suggests that generating hypotheses with CTC provides stable monotonic alignments under acoustic variability, while attention-based rescoring helps refine these hypotheses using stronger contextual modeling. Among the scratch-trained baselines, Conformer remains the most competitive architecture under all decoding settings.

\subsection{Pre-trained models}
Table~\ref{tab:whisper_results} shows that pre-trained systems provide a clear advantage over models trained from scratch on WildElder, reaffirming the value of large-scale pre-training for robust ASR. However, recognition on this dataset remains challenging, likely due to ageing-related voice characteristics, spontaneous speaking styles, and in-the-wild recording conditions. Across all configurations, fine-tuning on WildElder consistently improves performance over zero-shot inference, demonstrating that domain adaptation is essential for this acoustic domain.

Among the evaluated systems, Conformer-WenetSpeech delivers the strongest overall performance in both zero-shot and fine-tuned settings, which is consistent with its Mandarin-centric pre-training and closer domain match. The Whisper family exhibits a clear scaling trend: larger variants achieve better recognition accuracy, and all variants benefit from fine-tuning. The relative gains tend to be more pronounced for smaller models, while larger models still improve but show smaller marginal gains, suggesting diminishing returns with scale once strong pre-training is in place. Overall, these results highlight that multilingual pre-training offers robust cross-domain initialization, yet task- and domain-specific adaptation remains necessary to handle the variability and demographic characteristics present in elderly speech.

\begin{table}[t]
\centering
\caption{Conformer-WenetSpeech fine-tuning results by gender and age. 
\#Sent: number of sentences; N: total reference characters; S/N, D/N, I/N: substitution, deletion, and insertion rates (\%); CER: character error rate (\%).}
\vspace{2pt}
\label{tab:error-stats}
\begin{tabular}{l c c c c c c}
\toprule
\textbf{Cate.} & \textbf{\#Sent} & \textbf{N} & \textbf{S/N} & \textbf{D/N} & \textbf{I/N} & \textbf{CER} \\
\midrule

Female & 1282 & 25985 & 7.12 & 2.13 & 1.19 & 10.44 \\
Male   & 1118 & 24151 & 10.77 & 3.18 & 2.94 & 16.89 \\
[2pt]\hdashline[0.8pt/2pt]\addlinespace[2pt]

70--75 & 1477 & 33715 & 8.06 & 2.26 & 2.53 & 12.85 \\
75--80 & 383  & 6288  & 11.50  & 3.67 & 1.00  & 16.17 \\
80--85 & 380  & 8264  & 8.75  & 3.06 & 0.96  & 12.77 \\
85--90 & 130  & 1615  & 14.49  & 4.33  & 12.38  & 20.06 \\
90--95 & 30   & 254   & 19.69   & 1.97   & 2.76   & 24.41 \\
\bottomrule
\end{tabular}
\end{table}

\subsection{Analysis by Gender and Age}

We evaluate the fine-tuned Conformer-WenetSpeech model across gender and age groups to understand how speaker characteristics affect recognition accuracy in Table~\ref{tab:error-stats}. The system performs noticeably better on female speakers, reaching a character error rate of about 10.4\%, while the rate for male speakers is considerably higher at roughly 16.9\%. This gap suggests that the acoustic properties of elderly male speech introduce greater variability and pose additional challenges.

When broken down by age, the CER gradually increases with advancing age, indicating that the more pronounced vocal changes and reduced speech clarity in older participants adversely affect recognition. While performance remains relatively stable from ages 70 to 85, there is a marked degradation beyond 85, with the highest error observed in the 90 to 95 age range. These findings highlight the importance of targeted data collection and model adaptation to address the specific acoustic properties of very elderly speakers.

\section{Conclusion}

We presented WildElder, a Chinese elderly speech dataset collected from online videos and carefully curated with fine-grained manual annotations. The dataset provides broad demographic coverage, diverse acoustic conditions, and rich linguistic content, making it a realistic and challenging benchmark. Experimental results using both models trained from scratch and large-scale pre-trained systems show that elderly speech remains difficult, but fine-tuning substantially improves performance. WildElder offers a valuable resource for advancing elderly-oriented speech technologies and sets the stage for future research on robust modeling, adaptation, and inclusive human–computer interaction.

\section{Generative AI Use Disclosure}
Generative AI tools were used solely to edit and polish the manuscript, such as grammar correction and language refinement. All core content, including ideas, experiments, analyses, and conclusions, was developed entirely by the authors. The authors take full responsibility for the work and the contents of this paper.

\section{Acknowledgements}
This work was supported by NSF China (Grant No.62271270).

\bibliographystyle{IEEEtran}
\bibliography{mybib}

\end{document}